\documentclass[runningheads]{llncs}
\usepackage{llncsdoc}
\usepackage{graphicx}
\usepackage{color,soul}
\usepackage{amsmath}
\usepackage{xcolor}

\begin{document}
\title{A framework to construct a longitudinal DW-MRI infant atlas based on mixed effects modeling of dODF coefficients}
%
% \author{Heejong Kim\inst{1}\orcidID{0000-0002-9871-9755} \and
% Martin Styner\inst{2,3}\orcidID{0000-0002-8747-5118} \and
% Joseph Piven\inst{2}\and
% Guido Gerig\inst{1}\orcidID{0000-0002-9547-6233}} 
\author{Heejong Kim\inst{1} \and
Martin Styner\inst{2,3}\and
Joseph Piven\inst{2}\and
Guido Gerig\inst{1}} 
\authorrunning{H. Kim et al.}
\institute{Department of Computer Science and Engineering, New York University, NY, USA \and Department of Psychiatry, University of North Carolina, Chapel Hill, NC, USA \and Department of Computer Science, University of North Carolina, Chapel Hill, NC, USA}
\maketitle              
\setcounter{footnote}{0}
\begin{abstract}
Building of atlases plays a crucial role in the analysis of brain images. In scenarios where early growth, aging or disease trajectories are of key importance, longitudinal atlases\index{longitudinal atlas} become necessary as references, most often created from cross-sectional data. New opportunities will be offered by creating longitudinal brain atlases from longitudinal subject-specific image data, where explicit modeling of subject's variability in slope and intercept leads to a more robust estimation of average trajectories but also to estimates of confidence bounds. This work focuses on a framework to build a continuous 4D atlas from longitudinal high angular resolution diffusion images (HARDI) where, unlike atlases of derived scalar diffusion indices such as FA, statistics on dODFs is preserved. Multi-scalar images obtained from DW images are used for geometric alignment, and linear mixed-effects modeling from longitudinal diffusion orientation distribution functions (dODF) leads to estimation of continuous dODF changes. The proposed method is applied to a longitudinal dataset of HARDI images from healthy developing infants in the age range of 3 to 36 months. Verification of mixed-effects modeling is obtained by voxel-wise goodness of fit calculations. To demonstrate the potential of our method, we display changes of longitudinal atlas\index{longitudinal atlas} using dODF and derived generalized fractional anisotropy (GFA) of dODF. We also investigate white matter maturation patterns in genu, body, and splenium of the corpus callosum. The framework can be used to build an average dODF atlas from HARDI data and to derive subject-specific and population-based longitudinal change trajectories.

\end{abstract}
\section{Introduction}
Statistical brain atlases have become important models to provide standards to measure structural variations in anatomy, to define a common coordinate system for spatial correspondence between subjects, and as priors to segment brain structures. As diffusion weighted (DW) imaging yields micro-structural information, several studies have been proposed to construct DW atlases \cite{zhang2007unbiased,bouix2010building,yeh2011ntu,du2014geodesic,yang2017robust,kim2017graph,pietsch2018framework}. Atlas building\index{Atlas building} in DW imaging is challenging in that we need to consider diffusivity represented by multiple directional volumes rather than scalar images. Previous studies used diffusion tensor modeling \cite{zhang2007unbiased} and high order diffusion distribution for high angular resolution diffusion imaging (HARDI) data \cite{bouix2010building,yeh2011ntu,du2014geodesic,yang2017robust} to construct atlases using DW images. In a longitudinal DW atlas based on subject-specific longitudinal data, atlas building\index{atlas building} is more challenging as one faces temporal changes in addition to structural variability of sets of directional volumes. There are recently published works to build atlases with longitudinal HARDI images using fractional anisotropy (FA) maps and patch fusion to provide an early brain development atlas of infants \cite{kim2017graph} or considering multi-tissue time- and orientation-resolved group average atlases \cite{pietsch2018framework}. These studies, however, focus on building a template which reflects temporal differences of different age groups but not on building a fully continuous longitudinal atlas\index{longitudinal atlas}. 
\\\indent In this article, we introduce a new framework to construct a continuous longitudinal DW infant atlas based on linear mixed effects (LME) modeling of diffusion orientation distribution function (dODF) coefficients. The framework starts with building a multivariate HARDI template using FA and baseline images. We transform HARDI series to the template space with a reorientation step \cite{yap2012spatial}. Spherical harmonics (SH) coefficients are used to represent dODFs and for LME modeling of HARDI image time series. Considering repeated image data from longitudinal subject image series, we need to take into account the inherent correlation of repeated data and possibly unbalanced image time points which favor the use of mixed effects models. We applied our framework to a longitudinal dataset of HARDI images from healthy developing infants with ages between 3 to 36 months. We show the resulting longitudinal atlas\index{longitudinal atlas} with the time changes of dODF and its derived generalized fractional anisotropy (GFA). The following sections discuss our framework, experiments, and results.
\section{Method}
\begin{figure}[htb!]
\begin{center}
\includegraphics[width=0.9\textwidth]{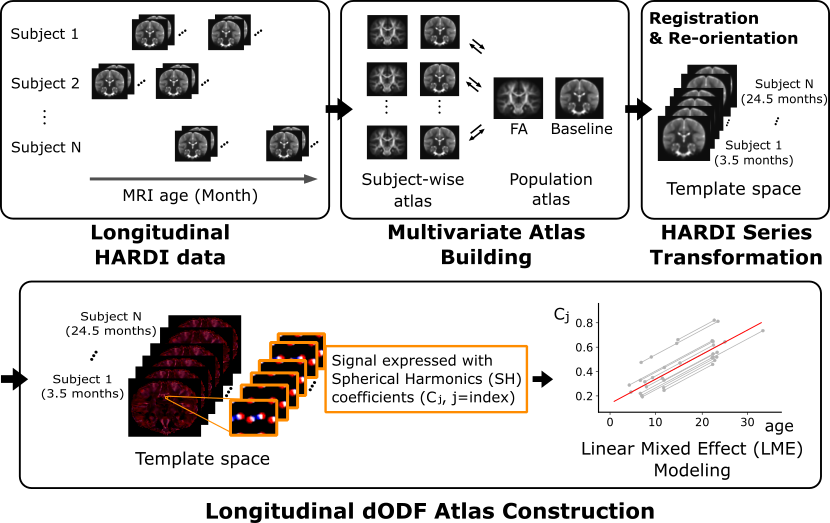}
\end{center}
\caption{Overview of longitudinal DW-MRI infant atlas building\index{atlas building} based on linear mixed effects modeling of diffusion orientation distribution function (dODF) coefficients.}
\label{figOverview}
\end{figure}
%
%\noindent\textbf{Multivariate Atlas Building}\quad
\subsection{Multivariate Atlas Building}
We use unbiased atlas building\index{atlas building} to create an anatomical average of the given population and time points. Construction of multivariate atlas is preferable in that the optimization satisfies both shape and appearance. We employed the symmetric group-wise normalization algorithm, which is a part of the open source toolkit Advanced Normalization Tools \cite{avants2011reproducible}. Given $N$ sets of the $K$ multi-modality images, $\mathbf{I} = \{I_{1},...,I_{K}\}$, multivariate template construction computes the set of diffeomorphic transforms, $\{(\phi_1,\phi_1^{-1}),...,(\phi_N,\phi_N^{-1})\}$, the optimal multivariate template, $\mathbf{J} = \{J_{1},...,J_{K}\}$, minimizing the cost function:
\begin{equation}
    {\textstyle\sum}_{n=1}^{N}\Big[D(\psi(\mathbf{x}),\phi_1^n(\mathbf{x},1))
    + {\textstyle\sum}_{k=1}^{K}\Pi_{k}(I^n_{k}, J_k(\phi_{n}^{-1}(\mathbf{x},1)))\Big]
\label{eq:atlas}
\end{equation}
where $D$ is the diffeomorphic shape distance, $D(\phi(\mathbf{x},0),\phi(\mathbf{x},1)) = \int_0^1 ||v(\mathbf{x}, t)||_L dt $ which depends on the linear operator, $L$, $v$ is the velocity field $v(\phi(\mathbf{x}, t)) = d\phi(\mathbf{x},t)/dt, \phi(\mathbf{x}, 0) = \mathbf{x}$, and $\Pi_k$ is the similarity metric. We use baseline images and FA maps to construct the multivariate atlas so that $K=2$ in Eq.~\ref{eq:atlas}. The normalized cross-correlation similarity metric has been suggested previously for multi-modality registration problems.
%
% \bigskip
%\noindent\textbf{Registration of HARDI series to Atlas}\quad
\subsection{Registration of HARDI series to Atlas}
Diffeomorphic transforms from the multivariate atlas building\index{atlas building} are used to transform HARDI series to a common coordinate system. In HARDI, we need an additional step which correctly reorients the diffusion profile for angular alignment in addition to transforming image series for structural alignment. This reorientation step is applied at each voxel in each HARDI series using a DW spatial warping algorithm \cite{yap2012spatial} as follows: Weighted diffusion basis functions are (1) computed by decomposition; (2) reoriented based on a local affine transformation; and (3) recomposed to the reoriented HARDI signal. The main advantage of using this warping algorithm is that it works directly on the signal so that we can make use of the reoriented HARDI signal for further processing. 
% \bigskip
%\noindent\textbf{Longitudinal Atlas Building from dODFs} \quad
\subsection{Longitudinal Atlas Building from dODFs}
We calculate dODFs for each voxel from HARDI data sets from all subjects and time points. HARDI signals can be represented by functions on the unit sphere where we can express the signal $S(\theta, \phi)=\sum_{l=0}^L\sum_{m=-l}^{l}c_{l,m}Y_{l,m}$
The basis functions $Y_l^m$ are given by:
\begin{equation}
    Y_l^m( \theta , \phi )  \sqrt{\frac{(2l+1)(l-m)!}{4\pi(l+m)!} } P_l^m(\cos \theta ) e^{im \phi }
\end{equation}
where $P_l^m$ is the associated Legendre polynomial, $l$ is the order, and $m \in [-l, l]$ is a phase factor. We calculated dODFs using a symmetric, real, orthonormal SH basis based on an analytical Q-ball imaging reconstruction method \cite{descoteaux2007regularized}. With the modified SH basis, we can write the set of equations as a linear system to solve for the coefficients $c_j$, where $j$ represents the index of coefficients, $j \in [1,(l^2+l+2)/2+m]$.
\\\indent Continuous longitudinal modeling is obtained by applying LME modeling to the sets of SH coefficients of ODFs measured at discrete time points. The LME model is preferable over linear regression as it takes into account the variability between subjects, the correlation of repeated data and, unbalanced data points. For each SH coefficient value, we evaluated longitudinal trajectories. Subject-wise intercepts are considered to have random effects and the group-wise slope is considered as the fixed effect representing estimated trends. The model for SH coefficients for the population of subjects and repeated measures over time can be formulated as follows,
\begin{equation}
    c_j \sim X\beta + Z\alpha + \epsilon,
\label{eq:LME}
\end{equation}
where $\mathbf{\alpha} \sim \mathcal{N}(0,\delta^2I)$, $X=[1, t]$, $Z$ is a design matrix and $\epsilon$ is an error term. $\mathbf{\beta}$ is a fixed effects vector and $\mathbf{\alpha}$ is a random effects vector. In our work, $t$ is MRI age and subjects are random effects factors. The LME formulation Eq.~\ref{eq:LME}, therefore, estimates group-wise slopes and intercepts as an average of subject-wise slopes and intercepts in the groups. From the estimated slopes of $c_j$ coefficients, we calculate the group-wise dODF trend at each voxel which results in a continuous longitudinal dODF atlas. The subject-wise dODF trends from the random effects represent individual subject-specific variability. 
\section{Experiments and Results}
%\noindent\textbf{Subjects}\quad
\subsection{Subjects}
Image data is selected from a population of 3- to 36-month-old children scanned on 3-T Siemens TIM Trio scanners on four different sites as part of an ongoing autism infant imaging study (ACE-IBIS). In this paper, thirty-three preprocessed HARDI scans from healthy developing infants with scans at more than one time-point are included to build atlas. The preprocessing pipeline includes quality control and correction techniques which are \verb|DTIPrep|\footnote{\url{https://github.com/NIRALUser/DTIPrep}} and Q-space resampling for correction~\cite{elhabian2016compressive}. DWI datasets were acquired with FoV = $209 mm$, 76 transversal slices, voxel size = $2\times2\times2 mm^3$ voxel resolution, matrix size = $106 \times 106$, TR = $11100 ms$, TE = $103 ms$, one baseline image with zero b-value and $64$ directional DWI volumes sampled on the half sphere with b-value at $2000 s/mm^2$, with a total scan time of 12.5 min.
%
%\bigskip
%\noindent\textbf{Multivariate Atlas Building and HARDI Series Registration}\quad
\subsection{Multivariate Atlas Building and HARDI Series Registration} 
\begin{figure}[htb!]
\begin{center}
\includegraphics[width=0.95\textwidth]{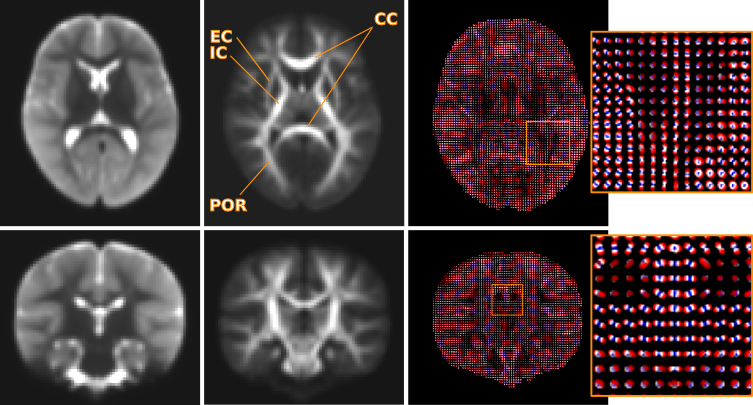}
\end{center}
\caption{Multivariate atlas construction result of baseline of DW image (Left) and fractional anisotropy (FA) (Middle). Visualization of dODFs calculated by averaging Spherical Harmonics (SH) coefficients across all HARDI series registered and reoriented into the template space (Right).}
\label{figMultiAtlas}
\end{figure}

\begin{figure}[htb!]
\begin{center}
\includegraphics[width=0.90\textwidth]{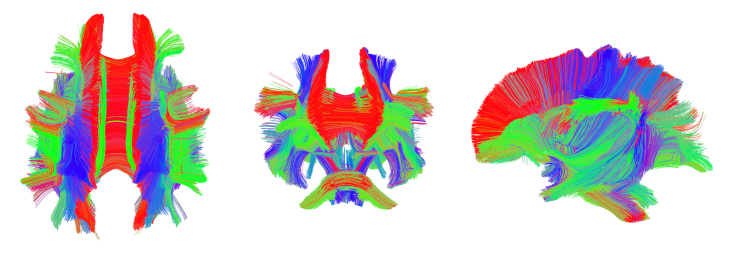}
\end{center}
\caption{Whole brain tractography from the average dODFs (Figure~\ref{figMultiAtlas}) in the template space: Axial (Left), coronal (Middle), sagittal (Right) view. We used generalized fractional anisotropy (GFA) from the dODF model to classify white matter. Deterministic fiber tracking with 30 degrees maximum angle threshold is used to obtain the tractography image. Tractography color shows the directions: red for left/right, blue for dorsal/ventral, and green for anterior/posterior.}
\label{figAtlasTractography}
\end{figure}

We use multivariate atlas building\index{atlas building} to obtain an anatomical average of the given population and time points. Fractional anisotropy (FA), the degree of anisotropy derived from the eigenvalues of diffusion tensor, has been used in building DW atlases for spatial alignment \cite{bouix2010building,geng2012quantitative,goodlett2009group}. Despite the advantage of representing locations of strong white matter tracts, FA maps do not fully represent boundaries of anatomical and fluid structures. There are different diffusivity indices representing different microstructure properties such as mean diffusivity (MD), radial diffusivity (RD), and axial diffusivity
(AD). The diffusivity indices, which are FA, MD, RD and AD maps, are calculated measures from eigenvalues of the diffusion tensor. The DW baseline image, which is a T2-weighted image, depicts structural properties not explained by FA maps. To decide which scalar values to be used for the multivariate atlas building\index{atlas building}, we compared a similarity between FA and the other scalar images by calculating normalized cross-correlation coefficient for all image pairs. Baseline and FA are the least correlated result in our infant dataset (Mean:$0.62$, std:$0.05$). The correlation coefficient values from combination pairs of MD, RD, AD, and baseline are higher than $0.95$ in average meaning that they represent similar structures. Thus, FA map and baseline were used to build multivariate atlas. 
\\\indent The atlas building\index{atlas building} approach includes two steps, building of subject-specific atlases, and of a population atlas. The initial subject-wise atlas building\index{atlas building} step copes with much lower within-subject variability as compared to across subject deformations. We compared our resulting atlas to an alternative atlas building\index{atlas building} without subject-wise atlas construction step. For each subject, the variance of normalized cross-correlation between the subject's scans and template was calculated. 12 out of 14 subjects showed lower variance in the atlas building\index{atlas building} with subject-wise step, justifying our choice. 
\\\indent Results of multivariate atlas building\index{atlas building} are shown in Figure~\ref{figMultiAtlas} with the resulting baseline atlas is in the first and FA atlas is in the middle column. The images in the third column illustrate the dODF visualization from averaged spherical harmonics (SH) coefficients of all infant HARDI images registered and reoriented to the template space. At each voxel, the average dODF is calculated from averaged coefficients. The known white matter tracts in the central part of brain including corpus callosum (CC), internal capsule (IC), external capsule (EC), and posterior optic radiation (POR) are clearly noticeable. We obtained whole brain tractography of the average dODF in the template space as additional visualization method (Figure~\ref{figAtlasTractography}). We used the deterministic tracking algorithm using \verb|Dipy|\footnote{\url{https://dipy.org/}} package (version 0.16.0.0). We set a maximum 30 degrees angle threshold and used generalized fractional anisotropy (GFA) from the dODF model to classify white matter. 
%\bigskip
%\noindent\textbf{Longitudinal dODF Atlas} \quad
\subsection{Longitudinal dODF Atlas}
\begin{figure}[htb!]
\begin{center}
\includegraphics[width=0.95\textwidth]{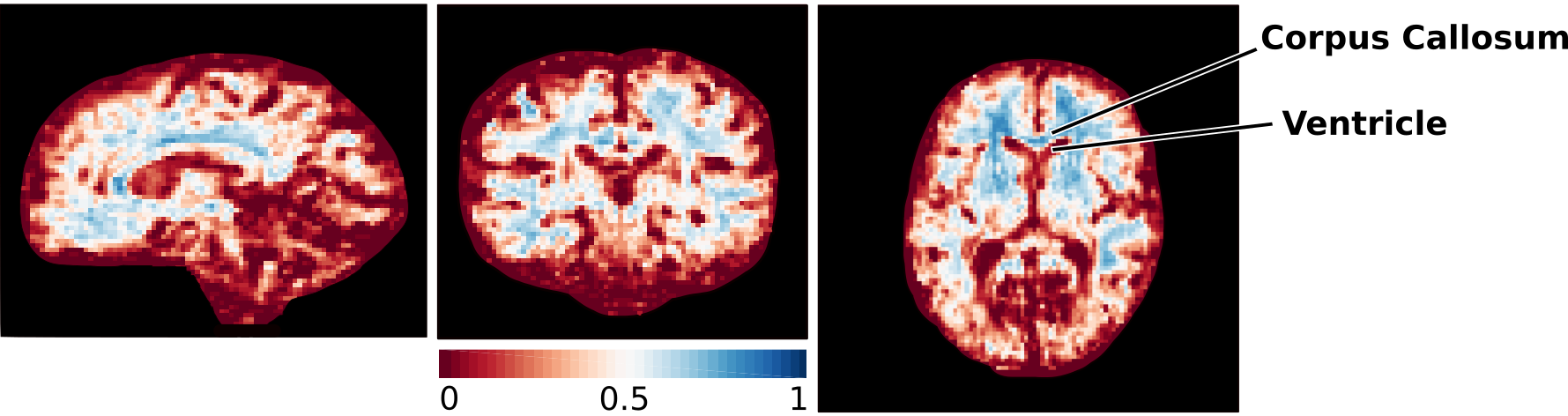}
\end{center}
\caption{Goodness of fit for the linear mixed effect (LME) result on the voxel-wise spherical harmonics (SH) coefficients. $R^2$ is calculated using the Frobenius norm for each voxel. Left: Sagittal view, Middle: Coronal, Right: Axial view}
\label{figRsquared}
\end{figure}

\begin{figure}[htb!]
\begin{center}
\includegraphics[width=1\textwidth]{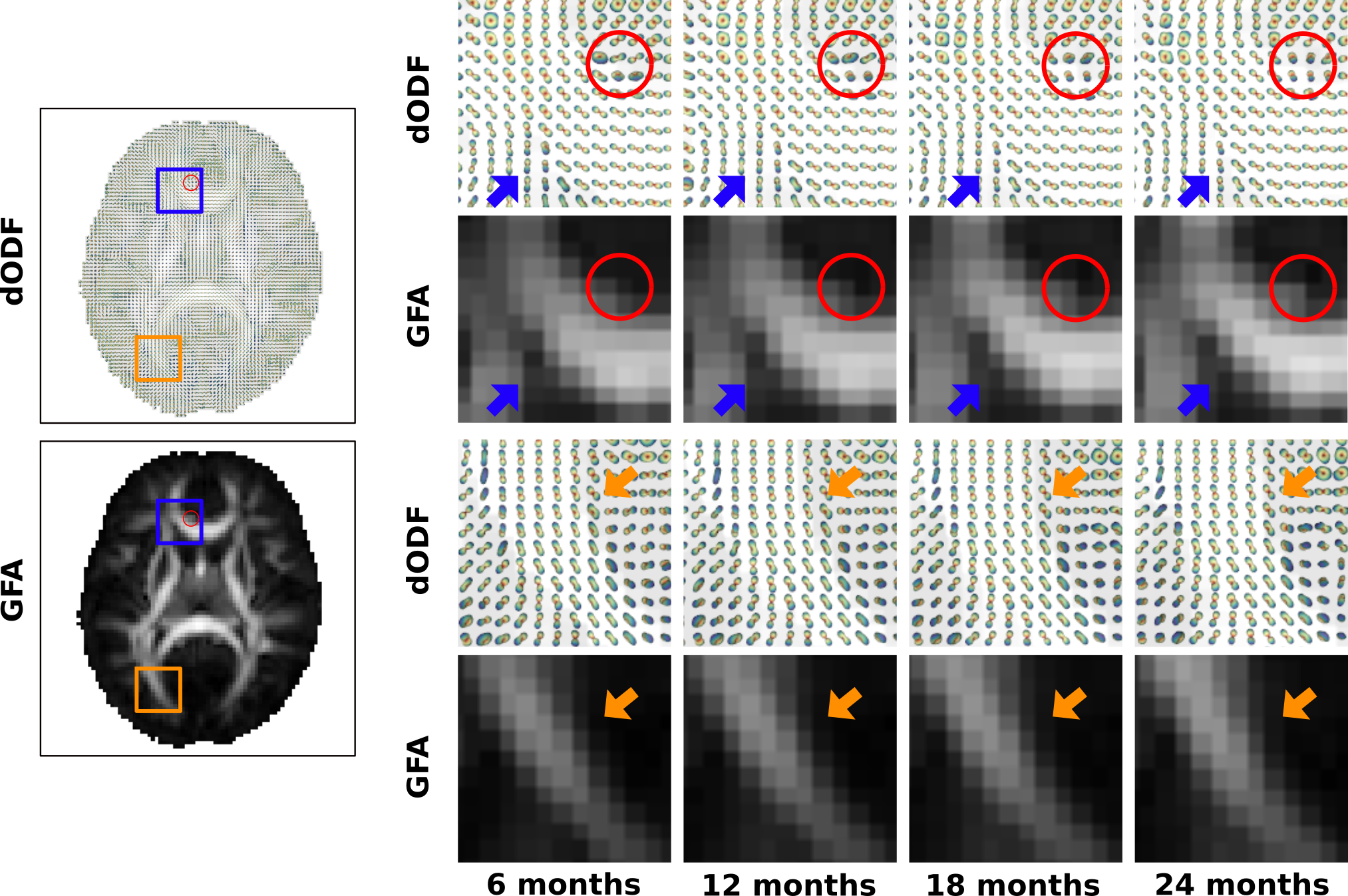}
\end{center}
\caption{Visualization of the 4D atlas result of longitudinal continuous diffusion orientation distribution functions (dODF) shown for cingulum (red circle), corpus callosum (blue arrow), posterior optic radiation (orange arrow). Each column on the right shows estimated dODFs and calculated generalized fractional anisotropy (GFA) at different time points.}
\label{figLongitudinal}
\end{figure}

To calculate a longitudinal dODF atlas, we first estimate ODFs from transformed HARDI images. The SH basis of order 6 ($l=6$), which makes 28 SH coefficients ($j=1,...,28$), with $\lambda=0.006$ has been selected following suggestions from Descoteaux's paper \cite{descoteaux2007regularized}. The order-6 terms consider up to 3 crossing directions in the orientation estimation. We adopted an LME modeling to obtain a continuous longitudinal dODF atlas. The model is applied to dODF coefficients for each voxel to estimate the group trends. To assess the goodness of fit, we calculated the $R^2$ on the voxel-wise SH coefficients $\mathbf{c}$ using the Frobenius norm (Figure~\ref{figRsquared}). $R^2$ will be near zero if the difference between estimated values and data is small due to the absence of longitudinal changes. For voxels located in regions having small diffusion signal changes such as cerebrospinal fluid, we get near zero $R^2$ values. Figure~\ref{figRsquared} shows the clear boundary of existing white matter structures, for example, corpus callosum and ventricle area, which can display the LME model has fitted well in voxels with white matter structures. We illustrate the estimated dODF and the generalized fractional anisotropy (GFA) of the resulting continuous longitudinal dODF atlas at different time points. We followed the GFA that Tuch proposed which can be interpreted as FA value but being able to be calculated from dODF \cite{tuch2004q}. We show the dODF and the GFA (Figure~\ref{figLongitudinal}) changes in the age range from 6 to 24 months. Figure~\ref{figLongitudinal} illustrates how the white matter structures change in the developing infant brains. The structure, for example, cingulum which is a red circle in the figure~\ref{figLongitudinal}, has a more clear directional shape in dODF as development continues. 
%
%\bigskip
%\noindent\textbf{Evaluation based on Longitudinal GFA Changes}\quad
\subsection{Evaluation based on Longitudinal GFA Changes}
\begin{figure}[htb!]
\begin{center}
\includegraphics[width=1\textwidth]{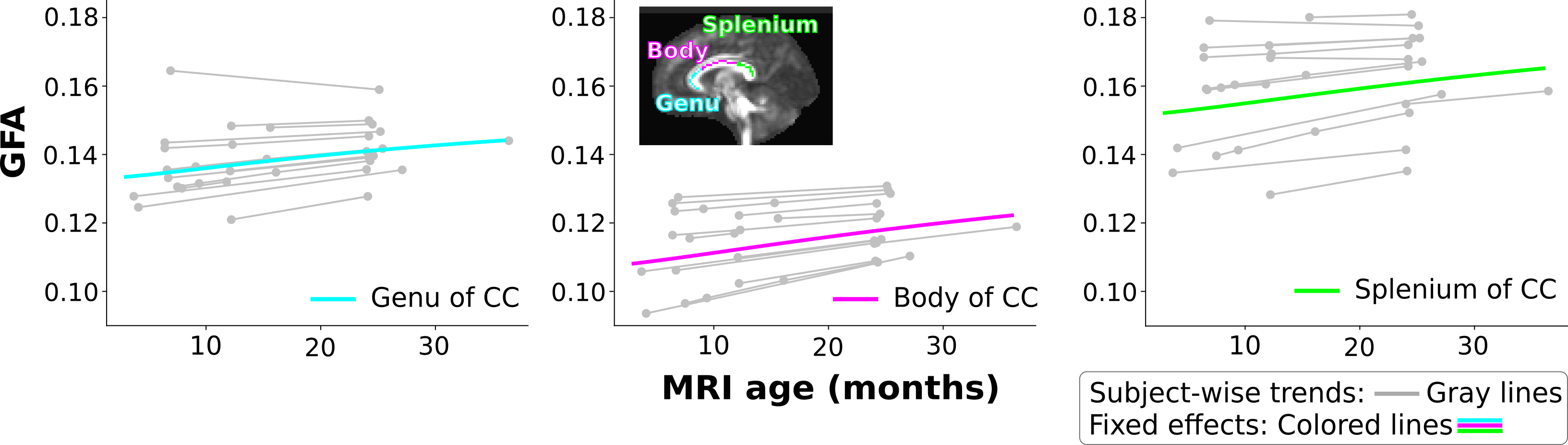}
\end{center}
\caption{Generalized fractional anisotropy (GFA) calculated from the longitudinal diffusion orientation distribution functions (dODF) infant atlas for genu, body, and splenium of the corpus callosum.}
\label{figROIGFA}
\end{figure}

In order to show the potential use of a longitudinal dODF atlas, we evaluate the GFA values calculated from dODFs in the corpus callosum (CC). In figure~\ref{figROIGFA}, the colored lines describe the population trends and the gray lines are the subject-wise trends. We analyzed three parts of the CC, genu, body, and splenium. The positive slope plots imply that the diffusivity of those area becomes more anisotropic as the brain develops. In addition, the splenium of CC starts with higher GFA which may indicate that the maturation of the splenium begins earlier compared to the body and the genu. This result is in line with the findings in a previous white matter tract-based study on DTI atlas, reporting that in neonates, the splenium shows highest FA, followed by genu and then the body of the CC~\cite{geng2012quantitative}. However, the study showed that changes in the first year were larger than the second year which could not be found with our framework since we assumed linear changes in dODF coefficients. Figure~\ref{figROIGFA} also illustrates the subject-specific variability in the GFA changes (gray lines). Developing statistics for individual trends to express confidence bounds will be subject of future work to develop methods such as age prediction will be our future work. 
\section{Discussion and Conclusion}
We introduce a framework to build a continuous longitudinal HARDI brain atlas based on statistics of dODF SH coefficients, applied to longitudinal data of healthy developing infants. The framework is generic and can be applied to any longitudinal study using DWI data, for example aging, disease progression or monitoring of therapeutic outcome. New concepts presented here are the use of longitudinal data for 4D atlas building\index{atlas building} which allowed the use of mixed effects modeling versus conventional regression, and continuous temporal modeling of dODFs from HARDI data, resulting in a 4D HARDI atlas where common scalar indices and variabilities can be derived. Geometric variability across subjects and age is normalized by unbiased atlas building\index{atlas building}, here using multi-modal image data to provide correspondences at anatomical boundaries as well as within interior white matter. We model continuous longitudinal dODFs by mixed effects modeling from longitudinal HARDI data, where fixed effects represent average and random effects variability as a function of age, a concept not shown before. Using the new framework, we will be able to not only to build average dODF atlases and derived scalar indices such as GFA at different time points, but also estimates of confidence intervals and thus variability as a function of time. Feasibility is shown with visualization of dODFs and derived GFA maps at multiple time points, demonstrating the ability to model maturation trajectories. Here, we could compare dODF atlas building\index{atlas building} and comparison of GFA with atlas-building directly from GFA maps, but the latter would only represent an atlas for one derived indices versus the rich set of measures to be derived from ODFs, and would lack the capability to apply fiber tractography from the atlas. The voxel-wise goodness of fit serves as validation of longitudinal modeling. As a potential use of the 4D atlas, we demonstrate GFA analysis of genu, body, and splenium of the corpus callosum which replicates previous findings~\cite{geng2012quantitative} based on DTI. 
\\\indent The presented work still has limitations. First, the unbalanced age points and the relatively small sample size may not allow to fully explain development between 3- to 36-month-old infants. This may lead to the result that the developmental effect is smaller than the inter-subject variability. Second, we assumed a linear change in time. The linear model may not be able to fully reflect non-linear age-dependent changes. 
\\\indent In future work, we will quantify the expected robustness of LME modeling versus cross-sectional atlas building\index{atlas building} and thus compare our concept to previously proposed atlas construction schemes. Future efforts will test improved geometric normalization via longitudinal regression, explore alternative mixed effects models for dODF coefficient modeling and further direct modeling method for diffusion MRI signals, examine higher order or nonlinear temporal models, and include a statistical framework for estimation of confidence bounds~\cite{sadeghi2013multivariate}. We will also extend the framework for longitudinal tract-based analysis of HARDI image data following concepts based on DTI~\cite{goodlett2009group}.
\section*{Acknowledgements}
This work was supported by the NIH grants R01-HD055741-12, 1R01HD089390-01A1, 1R01DA038215-01A1 and 1R01HD088125‐01A1. We are thankful for research discussions with Dr. Ragini Verma.
\bibliographystyle{splncs04}
\bibliography{miccai2019}

\noindent\textbf{Conflict of Interest Statement}: The authors declare that there are no conflicts or commercial interest related to this article 
\end{document}